# Impact of Low-OSNR Operation on the Performance of Advanced Coherent Optical Transmission Systems

P. Poggiolini[(1)], A. Carena[(1)], Y. Jiang[(1)], G. Bosco[(1)], V. Curri[(1)], F. Forghieri[(2)]

[(1)] Politecnico di Torino, DET, corso Duca degli Abruzzi, 24,10129 Torino, Italy, poggiolini@polito.it
[(2)] Cisco Photonics Italy srl, via Philips 12, 20900 Monza, fforghie@cisco.com

**Abstract** *We find evidence that low-OSNR operation causes substantial penalty on system maximum reach due to non-linearity generated by ASE noise and due to signal-power conversion into non-linear noise. Neglecting these effects may lead to a quite substantial performance prediction error.*

**Introduction**

The recent progress in coherent optical transmission technologies, together with major advances in forward error-correcting codes (FECs), has made it possible to carry out transmission at very low OSNRs. With PM-BPSK and PM-QPSK, at a FEC threshold $2.7 \cdot 10^{-2}$ [1], the corresponding values of the operating OSNRs (over a bandwidth equal to the symbol rate) are just 2.8 and 5.8 dB, respectively. Future FECs, possibly reaching $4.3 \cdot 10^{-2}$ [2], would further bring these OSNRs to as low as 1.7 and 4.7 dB, respectively. These low-OSNR scenarios are of particular interest for future ultra-long-haul systems that are expected to aim at 15,000 km (or more) such as in recently-announced trans-polar links [3].

In these conditions, ASE noise power is no longer small vs. signal power, at least in the last spans of the links. As a result, ASE impact on non-linear effects (or non-linear-interference, NLI) generation may become substantial.

In addition, since the non-linear process converts signal power into NLI noise, a depletion of signal power takes place too. This effect has been typically neglected in analytical performance predictions, but at low OSNRs it may become quite substantial.

In this paper we present the results of an investigation carried out on a 15-channel WDM PM-QPSK system at 32 GBaud. We compare high-accuracy simulations and analytical predictions, the latter performed using a recently proposed improved-accuracy variant of the GN model (the EGN-model [4], based on an extension and generalization of [5]).

Our bottom-line finding is that in current and future low-OSNR systems the impact of ASE-noise generated NLI and non-linear signal power depletion may be quite substantial. These findings have various implications on simulation techniques, NLI analytical modeling, system design criteria and performance limitations, which are discussed in the last section of the paper.

**System simulation set-up**

We focus on a 15 channel WDM PM-QPSK system, operating at 32 GBaud, with raised-cosine spectra (roll-off 0.05). The channel spacing is 33.6 GHz. The fiber is non-zero dispersion-shifted (NZDSF) with $\alpha$=0.22 dB/km, $D$=3.8 ps/(nm·km), $\gamma$=1.5 1/(W·km), with uniform span length of 120 km. We choose to use a relatively high-non-linearity fiber with long spans because simulations using SMF of PSCF with short spans would have been prohibitively time-consuming. However, we conjecture that the nature of the effects studied here is such that this choice does not affect the general findings, while keeping the computational effort manageable.

Lumped amplification is assumed, with ASE noise figure NF=5dB. The EDFA gain exactly equals the inverse of the nominal loss of the span. ASE noise is either loaded entirely at the receiver ("Rx-loading") or it is injected at each EDFA ("inline"). The photo-detected signals are first low-pass filtered through a 5-pole Bessel-type filter with bandwidth equal to half the symbol rate. Then they are sampled at two samples per symbol and sent to a chromatic-dispersion (CD) compensation stage. A 51-tap LMS equalizer follows, which first operates in training mode and then in decision directed mode. Each channel has its own 4 different PRBSs (degree 16). BER is estimated over 131˙072 symbols. To completely characterize the system for maximum reach, a full Rx is placed after each span. The launch power is stepped in intervals of 0.5 dB.

**Impact of ASE-generated NLI**

Fig.1 shows comprehensive reach results, for target BERs ranging from $3 \cdot 10^{-4}$ to $5 \cdot 10^{-2}$. The blue square markers represent simulations with Rx-loading, where ASE noise does not contribute to NLI generation. The red circles represent simulations with inline ASE, where ASE does contribute to NLI generation. At low BERs the two sets of markers are perfectly superimposed. As BERs go up, the markers start to split, notably above $10^{-2}$. The marker gap is shown in Fig.2, in dB, vs. BER. At the highest BER value ($5 \cdot 10^{-2}$) the gap is about 0.4 dB, or 10% of the maximum reach.

Rigorous modeling of this effect is possible in principle but it may be quite complex. However, a simple coarse model can be obtained as follows. We use the non-linear OSNR to predict performance:

$$\text{OSNR}_{\text{NL}} = \frac{P_{\text{ch}}}{P_{\text{ASE}} + P_{\text{NLI}}} \quad (1)$$

where $P_{\text{ASE}}$ and $P_{\text{NLI}}$ are the power of ASE and of NLI, respectively, over a bandwidth equal to the symbol rate. Assuming first, as it is usually done, that ASE noise contributes negligibly to NLI generation, we can write:

$$P_{\text{NLI}} = \eta_{\text{acc}}(N_{\text{span}}) \cdot P_{\text{ch}}^3 \quad (2)$$

where $\eta_{\text{acc}}$ is a non-linearity coefficient that depends on all link parameters other than launch power [6]. It also depends on $N_{\text{span}}$, as made evident in (2). Assuming the coarse approximation of *incoherent* accumulation of $P_{\text{NLI}}$ [6], then:

$$\eta_{\text{acc}}(N_{\text{span}}) \approx \eta \cdot N_{\text{span}} \quad (3)$$

where $\eta$ does not depend on $N_{\text{span}}$. Therefore:

$$P_{\text{NLI}} = \eta \cdot N_{\text{span}} \cdot P_{\text{ch}}^3 \quad (4)$$

We now want to account for the effect of inline ASE on NLI generation. We do so by summing, at each span, the power of the signal and the power of ASE noise accumulated up to there. Calling $P_{\text{ASE}}(n)$ the ASE power at the *n*-th span:

$$P_{\text{NLI}} = \eta \cdot \sum_{n=1}^{N_{\text{span}}} \left[ P_{\text{ch}} + P_{\text{ASE}}(n) \right]^3 \quad (5)$$

Substituting (5) into (1) we obtain a "corrected" OSNR$_{\text{NL}}$.

In Fig.2 we plot the gap between the maximum reach predicted using the OSNR$_{\text{NL}}$ with (4) and the corrected OSNR$_{\text{NL}}$ with (5). The curve is in rather good agreement with the gap found simulatively, suggesting that the $P_{\text{NLI}}$ model (5) captures the main mechanism causing the gap. Based on this model, we can predict what the gap would be when operating over the same link using PM-BPSK. The result is shown in Fig.2, too. The impact of ASE-generated NLI reaches 1 dB (20% max reach penalty) at BER= $5 \cdot 10^{-2}$.

**Impact of non-linear signal-power depletion**
To isolate the signal-power depletion effect from the previous one, we now focus on ASE Rx-loading only. We use again Eq.(1) and Eq.(2) to predict the maximum reach. This time, rather than coarse approximations such as (3), we use very accurate values of $\eta_{\text{acc}}(N_{\text{span}})$ found using the EGN model. The results are shown as thin dashed magenta lines in Fig.1. They start departing from Rx-loading simulations (blue squares) at BER $10^{-2}$ and the gap widens at higher BERs. At $5 \cdot 10^{-2}$ the effect is significant and amounts to 3 spans, close to 10% of the estimated max reach.

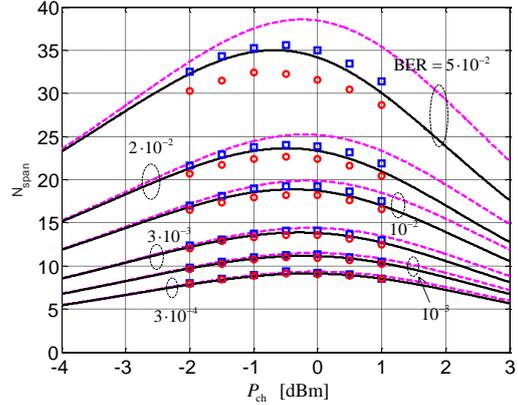

**Fig.1:** Reach in number of spans, vs. launch power per channel, for various values of target BER. System is quasi-Nyquist 15-channel PM-QPSK at 32 GBaud, NZDSF with span length 120 km. Blue squares: simulations with ASE noise loading at the Rx. Red circles: simulations with ASE noise injected inline at EDFAs. Dashed magenta line: EGN model with ASE noise loading at the Rx and OSNR as Eq. (1); black solid line: same, with OSNR as Eq. (6).

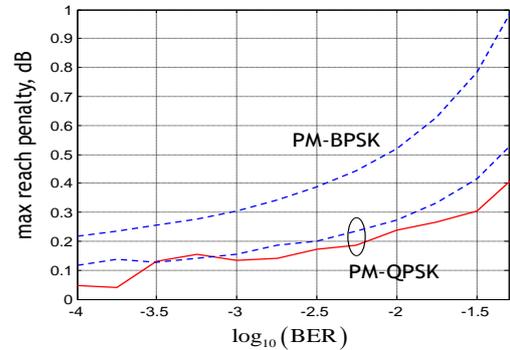

**Fig. 2:** Penalty due to the non-linear effect of inline ASE noise vs. ASE-noise loading at the Rx. Red solid line: simulations (gap between squares and circles of Fig. 1). Blue dashed line: coarse analytical model Eq. (1) and (3).

Since both the EGN model analytical results and the simulations are found *without* ASE noise in the line, this gap cannot be ascribed to ASE-related effects. We propose the following explanation. The non-linearity that is produced by the Kerr effect consists of converting part of the signal power into NLI noise. Assuming that EDFAs simply apply as much gain as fiber loss, this gradual depletion of the signal in favor of NLI noise is *not* compensated for by EDFA amplification. As a result, at the Rx the actual useful signal power is not $P_{\text{ch}}$ but approximately $(P_{\text{ch}} - P_{\text{NLI}})$, so that the actual OSNR$_{\text{NL}}$ is:

$$\text{OSNR}_{\text{NL}} = \frac{P_{\text{ch}} - P_{\text{NLI}}}{P_{\text{ASE}} + P_{\text{NLI}}} \quad (6)$$

Note that the expression for $P_{\text{NLI}}$ Eq. (2) is not affected because the amount of power generating NLI is, at each span start, the sum of the useful signal $(P_{\text{ch}} - P_{\text{NLI}})$ and of $P_{\text{NLI}}$ itself, i.e., still $P_{\text{ch}}$. When the corrected value of the OSNR Eq.(6) is used, in Fig.1 the black curves are found. They are in rather good agreement with the relevant simulations (the blue squares) and much closer to them than the EGN model calculations not using the correction (6) (i.e., the magenta lines).

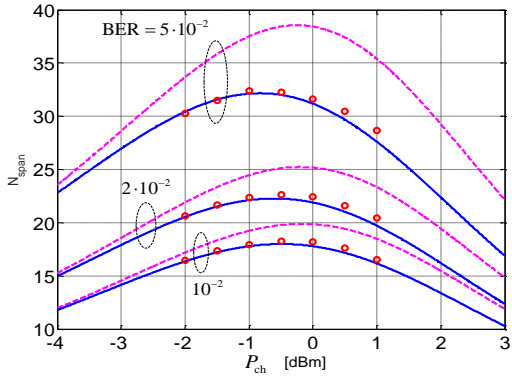

**Fig. 3:** Red Circles and dashed magenta lines: same as Fig. 1. Blue solid line: EGN model with Eqs. (6)-(8), accounting for both ASE-generated NLI and non-linear signal-power depletion.

**Combining corrections**

From Fig.1 one can see that the overall prediction error between the EGN-model magenta line and the red circles, while inexistent at BER $10^{-3}$, shoots up to 11% and almost 20% at BER $2 \cdot 10^{-2}$ and $5 \cdot 10^{-2}$, respectively. We try here obtain a more accurate analytical estimate of the maximum system reach at high BERs, based on the EGN model and the two corrections (5) and (6) together. As mentioned, exactly modeling the effect of inline ASE within the EGN model is ideally possible, but complex. We propose to approximate its effect by using an approximation similar to (5), namely:

$$P_{\text{NLI}} \approx \sum_{n=1}^{N_{\text{span}}} \eta_{\text{EGN}}(n) \cdot \left[ P_{\text{ch}} + P_{\text{ASE}}(n) \right]^3 \quad (7)$$

where $\eta_{\text{EGN}}(n)$ is the accurate value of $\eta$ found through the EGN model for the $n$-th span, no longer independent of $n$ as it was assumed in Eq. (5). This $P_{\text{NLI}}$ was used at the denominator of (6). At the numerator instead the value:

$$P_{\text{NLI,dep}} \approx \sum_{n=1}^{N_{\text{span}}} \eta_{\text{EGN}}(n) \cdot P_{\text{ch}}^3 \quad (8)$$

was subtracted from $P_{\text{ch}}$. The rationale here was that now part of NLI is produced depleting the signal and part depleting ASE. As a heuristic cirrection, the smaller value (8) is subtracted from the signal, instead of the total $P_{\text{NLI}}$ of (7). In Fig.3 the red circles are again simulations with inline ASE-noise and the magenta lines are Eq.(1) and (2). The blue solid lines are Eq. (6) with (7) and (8). The residual maximum reach prediction error is now very small.

**Discussion**

The trend towards the use of evermore powerful FECs and hence operation at lower OSNRs has been strong over the last several years and still continues. The results of this paper show that some of the expected advantages of such high-performance FECs may be thwarted by effects that are completely negligible at higher OSNRs, such ASE-generated NLI and non-linear signal-power depletion. We showed that the system max reach can turn out to be 15-20% smaller than expected with PM-QPSK, when the FEC threshold approaches $5 \cdot 10^{-2}$. The impact on PM-BPSK may be twice as much.

From the viewpoint of simulations, our findings show that, at very low target OSNRs, it is necessary to include inline injection of ASE noise to avoid substantial reach overestimation. From the standpoint of analytical modeling, it is necessary to deduct from the signal the amount of power that is transferred to NLI noise through the Kerr effect. A possible approximate way to do so is through the modified non-linear OSNR of Eq. (6). As for including in advanced NLI models the effect of inline ASE, we have proposed coarse approximations which capture the essence of the effect, while keeping the analytical complexity low. A rigorous solution to this problem is likely to be possible, but much higher complexity can be expected.

This work was supported by CISCO Systems within a SRA contract.